# Quantum Entanglement and Teleportation
Brent R. Yates
Physics 555B
Spring 2010

**Abstract**

Even Einstein has to be wrong sometimes. However, when Einstein was wrong he created a 70 year debate about the strange behavior of quantum mechanics. His debate helped prove topics such as the indeterminacy of particle states, quantum entanglement, and a rather clever use of quantum entanglement known as quantum teleportation.

**Introduction**

Quantum mechanics by nature is inherently a statistical field. A particle can never be predicted with absolute certainty, only with certain probabilities of being in a particular state. Even stranger is the idea that a particle is not in any given state but rather all states at once until it is observed. Once observed a particle is then permanently fix in the observed state. Einstein was convinced that quantum mechanics was simply a statistical approximation of a more concrete and precise description of the world.

In his attempts to "complete" quantum theory Einstein introduced several ideas such as local hidden variables (parameters that could not be directly seen, but affected and ultimately determined the state of a particle). He tried to use these hidden variables to solve the "problem" of entanglement and to prove that quantum particles indeed always had fixed values. Einstein was ultimately proven wrong, but he was "brilliantly wrong" (1). The disproval of Einstein's theories ultimately lead to concrete proof of phenomena such as entanglement, and entanglement was then exploited to allow the teleportation of quantum particles.

**Theoretical**

Entanglement

The basic idea of quantum entanglement is two or more "entangled" particles have direct influence on each other, even at a distance. The case of two particles is the simplest to describe. Assume each of the two particles can take a state 0 or 1 (often referred to as qubits in quantum computing). Due to entanglement, these two particles are in a superposition state

$$|\Psi\rangle_{12} = \frac{1}{\sqrt{2}}(|0\rangle_1|1\rangle_2 + e^{i\chi}|1\rangle_1|0\rangle_2)$$

where $\chi$ is a phase shift determined by the internal properties of the source (2). For further simplicity setting $\chi = 0$ reduces the superposition to $|\Psi\rangle_{12} = \frac{1}{\sqrt{2}}(|0\rangle_1|1\rangle_2 + |1\rangle_1|0\rangle_2)$. According to quantum mechanics neither of these particles carries a definite value until observed. As soon as a measurement is made on one (the outcome being completely random) the second particle immediately assumes a well defined quantum state orthogonal to the observed particle.

An example of entanglement is a pion decaying into an electron and a positron (3). The particles travel in opposite directions and have opposite spin (each is a spin ½ particle so one must be spin up and the other must be spin down to conserve the angular momentum of the spin

0 pion). Because of the nature of the emission of the electron and positron they are quantumly entangled. Assume the electron goes to Alice and the positron goes to Bob. Alice then does an experiment on her electron to determine its spin along the z-axis. The experiment has a 50% chance of finding the electron along the +z-axis and a 50% chance of finding the electron along the −z-axis. Assume Alice finds her electron to be in the +z-axis. According to quantum mechanics Alice's experiment has collapsed the wave function of the electron and it is now permanently oriented along the +z-axis. Because the two particles are entangled Alice automatically knows that Bob's positron must be along the −z-axis. This troubled Einstein greatly as according to his special theory of relativity nothing can travel faster than the speed of light, yet Alice *instantly* knows the orientation of Bob's positron. Another problem is apparently no communication takes place between Alice and Bob's particles. Einstein called this "spukhafte Fernwirkung" or spooky action at a distance (1).

EPR

Although Albert Einstein made major contributions to the development of quantum mechanics he did not believe in many of its predictions. One of the biggest problems Einstein had with quantum mechanics was its statistical nature. According to quantum mechanics a particle's state can never be fully known; at best quantum mechanics can give the probability of a particle being in a certain state. Einstein made a famous quote saying "God does not play with dice." Over the years Einstein had many debates with Niels Bohr over the implications of quantum mechanics. One of the most famous debates was a paper written by Einstein, Podolsky and Rosen. In this paper they argued what is now known as the EPR paradox.

The EPR paper proposed a Gedanken (German for mind) experiment which attempted to use the phenomenon of quantum entanglement to show that quantum mechanics was incomplete. They proposed a quantum system of two particles in which the position and momentum of each particle is not well defined, but the sum of the positions (center of mass) and the difference of their momenta (individual momenta in the center of mass system) are both precisely defined (2). According to quantum entanglement a measurement of the position or momentum of one particle immediately implies the precise position or momentum of the other particle, even when separated by an arbitrary distance. Einstein and his colleagues argued that since nothing can travel faster than the speed of light then the measurement of one particle cannot have any influence on the other. Therefore both the position and momentum of each particle are simultaneously well defined.

Einstein, Podolsky and Rosen introduced LHV (local hidden variables) and argued that without these quantum mechanics was incomplete (1). They proposed that there were certain hidden variables that quantum mechanics did not account for, and taking these variables into account a particle would no longer be in an undetermined state. Bohr on the other hand argued that the two particles in the EPR experiment are always part of one quantum system. A measurement on one particle changes the possible predictions of the whole system, and therefore the other particle as well. The idea of LHV was later proven wrong by John Bell in 1964.

Bell's Inequality

Bell started by considering a source emitting two photons in an entangled state
$$|\Phi^+\rangle_{12} = \frac{1}{\sqrt{2}}(|V\rangle_1|V\rangle_2 + |H\rangle_1|H\rangle_2)$$
where one is sent to Alice and the other to Bob. Each would perform polarization measurements by passing their photons through a beam splitter with two single photon detectors at either output of the beam splitter. Each photon has an equal probability of being measured vertically or horizontally polarized, and as before if Bob chooses the same basis he always gets the same results as Alice (2). Next he assumed the EPR paradox where Alice can predict Bob's results simply from her own. He also assumed locality – no physical influence can go from Alice's apparatus to Bob's apparatus, therefore Bob's result only depends on properties of his photon and apparatus. Alice and Bob can chose their detection bases at any oblique angles. Bell took three arbitrary angles: α, β, and γ. Bell's inequality states

$$N(1_\alpha, 1_\beta) \leq N(1_\alpha, 1_\gamma) + N(1_\beta, 0_\gamma)$$

where

$$N(1_\alpha, 1_\beta) = \frac{N_o}{2} \cos^2(\alpha - \beta)$$

gives the number of times Alice detects a photon at angle α and Bob detects a photon at angle β and $N_o$ is the total number of photons emitted by the source. By simply letting $(\alpha - \beta) = (\beta - \gamma) = 30°$ then the inequality is violated. This implies that at least one of Bell's assumptions conflicts with quantum mechanics. This is usually viewed as evidence that non-locality exists (2).

Bell States

For two particles each with orthogonal states $|0\rangle$ and $|1\rangle$ there exists four maximally entangled Bell states (2)

$$|\Psi^+\rangle_{12} = \frac{1}{\sqrt{2}}(|0\rangle_1|1\rangle_2 + |1\rangle_1|0\rangle_2)$$

$$|\Psi^-\rangle_{12} = \frac{1}{\sqrt{2}}(|0\rangle_1|1\rangle_2 - |1\rangle_1|0\rangle_2)$$

$$|\Phi^+\rangle_{12} = \frac{1}{\sqrt{2}}(|0\rangle_1|0\rangle_2 + |1\rangle_1|1\rangle_2)$$

$$|\Phi^-\rangle_{12} = \frac{1}{\sqrt{2}}(|0\rangle_1|0\rangle_2 - |1\rangle_1|1\rangle_2)$$

These are of great value in quantum computing because 2 qubits can be encoded by manipulating only 1 particle.

Assume Alice and Bob each have one particle of an entangled pair. For simplicity consider the entangled pair to be in the $|\Psi^+\rangle_{12}$ state. There are four possible unitary transformations Bob can perform on his particle (particle 2). The first is the identity transformation in which the state does not change. The second is state exchange in which $|0\rangle_2 \rightarrow |1\rangle_2$ and $|1\rangle_2 \rightarrow |0\rangle_2$ and the state becomes $|\Phi^+\rangle_{12}$. The third is a state-dependent phase shift where $|0\rangle_2$ and $|1\rangle_2$ differ by π and the state becomes $|\Psi^-\rangle_{12}$. The fourth transformation is applying a state exchange and a phase shift so the state becomes $|\Phi^-\rangle_{12}$. Alice can then read Bob's encoding by determining which Bell state the system is in after the transformations (2).

Teleportation

One of the most interesting phenomena to come from entanglement and the EPR paradox is the idea of quantum teleportation. In the classical world if Alice wanted to tell Bob how to construct something she had, say a table, she would simply have to examine the table and send Bob the proper schematics. This same task cannot be easily performed in the quantum world due to the no-cloning theorem, first proved by William Wootters and Wojchiech Zurek in 1982 (1).

According to quantum theory all particles exist in a superposition of states. For example a photon is both horizontally and vertically polarized at the same time until observed. The problem with making an exact copy of a quantum particle is once one parameter is measured the wave function collapses, the superposition is destroyed, and no further information can be obtained about the particle.

Suppose Alice has a single photon (called particle 1) in the state $|\Psi\rangle_1 = \alpha|V\rangle + \beta|H\rangle$ where α and β are complex amplitudes ($|\alpha|^2$ gives the probability of finding the photon vertically polarized, $|\beta|^2$ gives the probability of finding the photon horizontally polarized, and $|\alpha|^2 + |\beta|^2 = 1$). Alice wants to send this photon to Bob. To do so classically she would have to know both α and β, send these values to Bob, and then Bob would have to find a way to create a photon with these same values. However, as soon as Alice does an experiment to determine α she destroys the superposition of her photon, and it is now permanently polarized in the vertical direction. This would leave Alice with no possible way of measuring β for this same photon. Alice must use quantum teleportation to properly transfer her photon to Bob (Figure 1). The theoretical process is not very complicated and boils down to four basic steps (1). In step one Alice and Bob must share a pair of entangled particles (particles 2 and 3). They can use photons, or any other particles that can be entangled. Step two requires Alice to make a Bell state measurement (BSM) on her particle 1 and the entangled particle 2. This projects particles 1 and 2 onto one of the four Bell states. An example of this would be passing particles 1 and 2 through a beam splitter and observing whether they both come out at the same side (same polarization) or on opposite sides (different polarizations). She gains no information about her particle; she simply knows whether particles 1 and 2 are in the same or different states. The possible projections are

$$|\Psi\rangle_{123} = |\Psi\rangle_1 \otimes |\Psi\rangle_{23} =$$
$$\frac{1}{2}[|\Psi^-\rangle_{12}(-\alpha|V\rangle_3 - \beta|H\rangle_3)$$
$$+ |\Psi^+\rangle_{12}(-\alpha|V\rangle_3 + \beta|H\rangle_3)$$
$$+ |\Phi^-\rangle_{12}(\alpha|H\rangle_3 + \beta|V\rangle_3)$$
$$+ |\Phi^+\rangle_{12}(\alpha|H\rangle_3 - \beta|V\rangle_3)]$$

(2). Notice that for photons the Bell states $|\Psi^+\rangle_{12}$ and $|\Psi^-\rangle_{12}$ are degenerate as are $|\Phi^+\rangle_{12}$ and $|\Phi^-\rangle_{12}$. This is because the beam splitter simply tells if the particles coincide, not the phase between them. In step three she sends the results of this experiment to Bob via classical means (telephone, email, etc.). In the fourth step Bob uses this information to perform the proper unitary transform on his particle 3. As an example assume Alice's BSM experiment resulted in the Bell state $|\Phi^-\rangle_{12}$. This would leave Bob's particle 3 in the state $\alpha|H\rangle_3 + \beta|V\rangle_3$. This means Bob should perform the state exchange transformation ($|H\rangle_3 \rightarrow |V\rangle_3$ and $|V\rangle_3 \rightarrow |H\rangle_3$). He now has an exact replica of Alice's particle 1. It should be noted that when Alice makes the measurement in step two her original particle 1 loses its properties and becomes entangled with particle 2, while particle 3 is no longer entangled with particle 2 but is now in an orthogonal state to Alice's original particle. The destruction of $|\Psi\rangle_1$ is due to the no-cloning theorem.

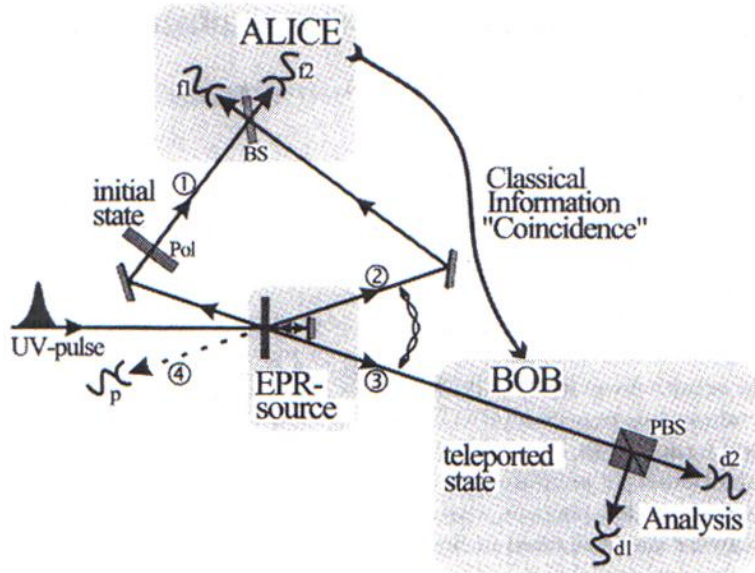

**Figure 1 – Teleportation Protocol (2)**

**Experimental**

Entanglement

Between 1981 and 1982 Alain Aspect conducted three experiments to test the quantum entanglement theory (1). Aspect's first experiment is pretty much identical to the Gedanken experiment proposed in the EPR paper. They excited a calcium atom which emitted two photons in opposite directions. Due to the nature of their emission the two photons were entangled. On either side of the source they placed a polarizer and a photon detector. They put the polarizers at various angles and counted the amount of times both detectors received a photon (Figure 2). The correlation they found was much greater than what LHV would predict, so much so that they claimed the probability of their observations happening by chance and not through entanglement was a one in $10^{36}$ (1). There were still a number of scientists who believed the LHV theory and argued problems with Aspect's experiment. For example, in the Gedanken experiment the detectors were perfect, but there is no such thing as a perfect detector in real life. If a detector reported a 0 they automatically assumed no photon was detected. This did not take into account the fact an imperfect detector might fail to detect a photon which was actually there. The argument of imperfect detectors prompted Aspect to conduct a second experiment.

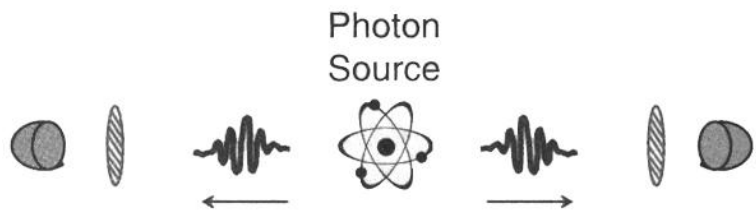

**Figure 2 – Aspect Experiment 1 (1)**

In his second experiment Aspect used the same source, but this time each photon would first pass through a beam splitter. Both beam splitters had a detector at each output, so Aspect had a total of two beam splitters and four detectors (Figure 3). Since a single photon has a 50% chance of coming out of either side of the beam splitter it should always be detected. If both detectors report a 0 then the one that the photon was sent to has failed and the current data point should be thrown out (1).

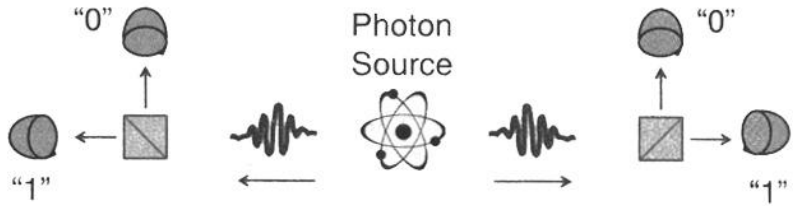

**Figure 3 – Aspect Experiment 2 (1)**

There was still some skepticism about Aspect's experiments. Some people believed that the detectors could possibly be sending signals to the source or the other photon. To resolve this they inserted an optical switcher to direct the each photon to one of their two detectors (Figure 4). Each detector also had a polarizer in front of it. The optical switcher changed every 10ns, but the photons took 40ns to go from the source to the detector. This meant the detector was not chosen until after the photon left the source.

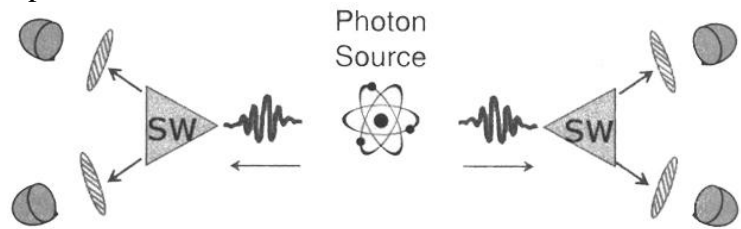

**Figure 4 – Aspect Experiment 3 (1)**

Teleportation

In 1997 Anton Zeilinger actually demonstrated quantum teleportation experimentally in Innsbruck, Austria (1) as illustrated in Figure 5. He and his team started by firing an ultraviolet laser into a special crystal. This crystal produced two entangled infrared photons (each with half the energy of the original UV photon). They did this twice to get two pairs of entangled photons. One pair was used as the entangled pair to send the information (particles 2 and 3). Of the other pair, one was sent through a polarizer to provide a state they wished to teleport (particle 1). Particle 4 was simply used as a trigger to let them know when to collect their data. Then they passed particles 1 and 2 through a beam splitter to perform a BSM, which projected particle 3 into a particular polarization. Unfortunately in this experiment they could only detect one of the four Bell states. The BSM measurement of particles 1 and 2 resulted in the Bell state $|\Psi^-\rangle_{12}$ which meant particle 3 was in the state $-\alpha|V\rangle_3 - \beta|H\rangle_3$. They then passed particle 3 through a polarizer which was properly rotated to only allow that state through, blocking all other orientations. Behind the polarizer was a photon detector. They performed this experiment several times, and each time they confirmed particle 3 was polarized at exactly the angle predicted. In 2004 Zeilinger repeated his experiment, this time teleporting a photon 600m across the Danube River in Vienna, Austria (1). They transferred their photons over an optical fiber to insure no scattering processes took place during the transfer which might have altered the state of their photons. Again they found that particle 3 was always in the state predicted by quantum teleportation theory.

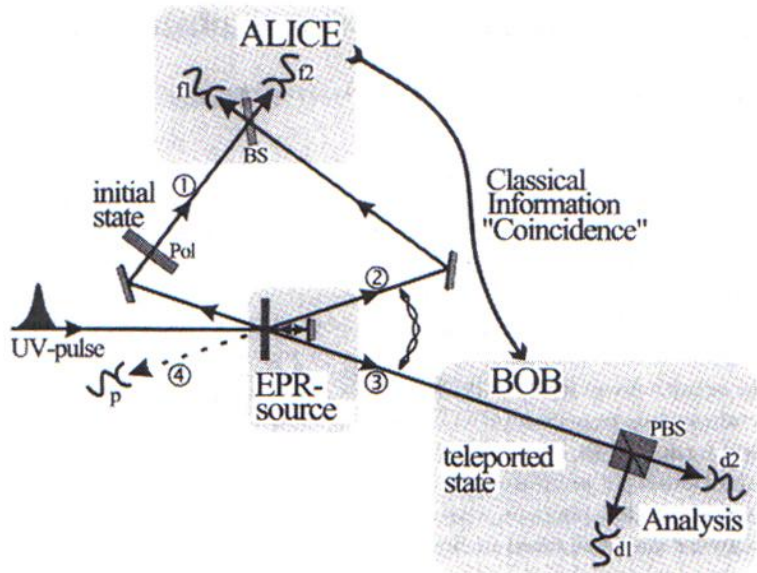
**Figure 5 – Zeilinger Teleportation Experiment (2)**

**Conclusion**

      Einstein set out to disprove quantum mechanics, or at least prove it was incomplete. Instead he started a philosophical debate which ultimately proved the validity of quantum mechanics. The ERP paradox combined with Bell's inequality proved non-locality which lead to the proof that quantum entanglement does exist. Quantum teleportation soon emerged as a result of quantum entanglement. Quantum entanglement and teleportation have many practical applications. The no-cloning theorem and the collapsing of superposition wave functions upon observation make it impossible to know all internal properties of a quantum particle. The only way to make a complete reproduction of a quantum particle is to use quantum teleportation. Quantum teleportation is also not limited to two particles. At the Niels Bohr Institute in 2006 an experiment was performed which teleported one trillion atoms (1). This may not be enough to teleport a human, but it shows that there seems to be no limit on the number of particles which can be teleported. Roger Penrose supports the idea that human consciousness is a quantum phenomenon. If Penrose and others who share this idea are correct then the only way to successfully transfer a person's consciousness is to use quantum teleportation to ensure every single particle in the brain is in the right state. Quantum entanglement and teleportation also have many applications in quantum computing and dense coding. Quantum entanglement has shown that two qubits can be manipulated by altering the state of only one. Entanglement can also help decrease computation time. If two qubits in a quantum computer are needed but are separated by some distance, then the computation could be carried out instantly via entanglement rather than waiting for the two particles to interact. Caltech also has a proposal for using quantum teleportation to create a quantum internet, linking two or more quantum computers together (1). Quantum entanglement and teleportation may seem like abstract theoretical ideas, but they have already been proven and will most likely have a profound effect in the not too distant future.